\documentclass[aps,prl,twocolumn,nopacs,superscriptaddress,floatfix]{revtex4-1} 

\usepackage{amsmath}
\usepackage{bbm}
\usepackage{epsfig}
\usepackage{color}
\usepackage{graphicx}
\usepackage{epstopdf}
\usepackage{amsfonts}
\usepackage{verbatim}

\newcommand{\id}{\mathbbm{1}}

\usepackage{times}

\newcommand{\cc}{{\mathbbm{C}}}
\newcommand{\rr}{{\mathbbm{R}}}
\newcommand{\Tr}{\mathrm{Tr}}

\newcommand{\EE}{\mathbbm{E}}
\newcommand{\poly}{\mathrm{poly}}
\newcommand{\polylog}{\mathrm{polylog}}
\newcommand{\PP}{\mathbbm{P}}

\newcommand{\eps}{\varepsilon}

\newtheorem{theorem}{Theorem}

\newtheorem{observation}[theorem]{Observation}
\newtheorem{requirement}[theorem]{Requirement}

\begin{document}

\title{Efficient measurement-based quantum computing with continuous-variable systems}

\author{M.\ Ohliger} 
\affiliation{Dahlem Center for Complex Quantum Systems, Freie Universit{\"a}t
Berlin, 14195 Berlin, Germany}
\affiliation{Institute for Physics and Astronomy, University of Potsdam, 14476 Potsdam, Germany}

\author{J.\ Eisert}
\affiliation{Dahlem Center for Complex Quantum Systems, Freie Universit{\"a}t 
Berlin, 14195 Berlin, Germany}

\begin{abstract}

	We present strictly efficient schemes for scalable measurement-based quantum computing using
	continuous-variable systems: These schemes are based on suitable non-Gaussian resource states,
	ones that can be prepared using interactions of light with matter systems or even purely optically.
	Merely Gaussian measurements such as optical homodyning as well as
	photon counting measurements are required, on individual sites. These schemes
	overcome limitations posed by Gaussian cluster states, which are known not to be 
	universal for quantum computations of unbounded length, unless one is willing to scale 
	the degree of squeezing with the total system size.
	We establish a framework derived from tensor networks and matrix product states
	with infinite physical dimension and finite auxiliary dimension general enough to provide a framework
	for such schemes. Since in the discussed schemes the logical encoding is finite-dimensional, tools of error
	correction are applicable. We also identify some further
	limitations for any continuous-variable computing scheme from which one can argue that no substantially easier ways of 
	continuous-variable measurement-based computing than the presented one can exist.
\end{abstract}
\maketitle

\section{Introduction}
\label{sec:intro}

To realize a quantum computer in the circuit model, it is crucial to have precise control over each of the carriers of quantum information. In addition to keeping the quantum state of the computer protected from unavoidable noise induced by the environment, it is necessary to implement suitable quantum gates, usually in the form of one and two qubits gates. The latter is of particular difficulty, especially when optical quantum systems are used because their interaction is either weak or merely induced by measurements. 

The paradigm of measurement-based quantum computing (MBQC) as pioneered by Raussendorf and Briegel \cite{clusterbriegel,clusterlong} and substantially generalized by Gross and Eisert \cite{prl,pra} allows to get rid of the necessity of performing unitary operations to implement a quantum circuit. Instead, the actual computation is performed by preparing a multipartite entangled state, the resource, in a first step followed by adaptively chosen local measurements on this resource. The important improvement stems from the fact that the resource is universal, i.e., it can be prepared independently of the algorithm one wants to perform. This means that the presumably difficult step, the one which involves entangling operations, can be performed off-line. This resource-preparation may also be probabilistic as it is possible to wait with the implementation of the algorithm until the resource is available. What is more, individual addressing in the final read out step is anyway also required in the circuit model, so this step does not add any further difficulty to the scheme.

Quantum computing based on continuous-variable (CV) quantum optical systems differs from the more conventional 
notion based on their discrete analogues in not making use of single photons as the carriers of quantum information \cite{cvreview1,cvreview2}. Instead, it relies to a large extent on Gaussian states and their manipulation. The notable advantage is that Gaussian states are easier to prepare in the laboratory, the corresponding interactions are often stronger and easier to accomplish, and some measurements, e.g.,  homodyne detection, can be performed with an efficiency substantially surpassing the one of single-photon measurements. However, quantum information protocols using Gaussian states only, Gaussian operations, and Gaussian measurements suffer from serious limitations as in this setting neither entanglement distillation \cite{ng1,ng2,ng3} nor error correction against Gaussian errors \cite{cerf} is possible. Also, since one can easily efficiently keep track of first and second moments, any Gaussian evolution of Gaussian states can be efficiently simulated on a classical computer \cite{efficientsimulation}, clearly ruling out as the possibility of universal quantum computing. 

MBQC based on Gaussian resource states has been extensively discussed in the literature \cite{gaussiancluster1,gaussiancluster2,gaussiancluster3,gaussiancluster4,gaussiancluster5} due to several features: They are a direct generalization of the well known qubit cluster-state to the continuous-variable regime, they allow for universal quantum computing with Gaussian measurements and a single non-Gaussian one, and they can be prepared with present day experimental techniques, as already demonstrated. However, if they are not formed out of idealized infinitely squeezed states (ones that are not contained in Hilbert space and would require infinite energy in preparation) but rather from physical states possessing only finite squeezing, they suffer from exponentially decaying localizable entanglement. This limitation, which applies to any Gaussian resource state irrespectively of the permitted class of measurements \cite{nogo}, implies that full error correction and the machinery of fault-tolerant quantum computing \cite{faulttolerantnielsen,faulttolerantraussendorf}, in parts yet to be developed for such continuous-variable systems, appears to be necessary even when both state preparation and measurements are perfect. Notably, those restrictions only apply when trying to perform a quantum computation of an \textit{unbounded} length. For any \textit{finite} length, there exists, for any required accuracy, a physical cluster state such that any quantum operation up to this length is possible with this accuracy. Because the widely assumed superior power of quantum computers compared to classical ones manifests itself most prominently in the scaling of the runtime with the problem size, the situation of a quantum computation with unbounded length is in the focus of attention in this work.

In addition to schemes based on Gaussian cluster states, two more classes of schemes relying on CV-quantum optics have been proposed. The first one is based on superpositions of coherent states which are called Schr{\"o}dinger cat states or, when they have very low amplitude, kitten states \cite{cats}. They can be, in approximation, created by subtracting photons from squeezed states and allow for universal quantum computing by passive operations such as beam-splitters and photon-counting detection only. However, they also suffer from quite severe problems: the probabilistic nature of the quantum operations stemming from the use of non-overlapping basis-elements seems to be the most fundamental one, giving rise to significant overheads. The second approach combines the advantageous features of both discrete and continuous-variable optics \cite{hybrid1,hybrid2}. In these schemes, quantum information is carried by qubits and single qubit operations are performed directly on them. Two-qubit operations, in contrast, are performed by letting both qubits interact with a strong CV-mode, the qubus. 

With this article, we pursue two different, but complementary goals: On the one hand, we aim at  clarifying the boundary between settings where CV-MBQC is possible and such situations where this is not the case. For this reason, we will develop a general framework capable of describing quantum computation in the measurement-based paradigm, regardless of the dimension of the carriers of quantum information. Within this picture, we can identify some serious limitations giving rise to challenges that have to be overcome. On the other hand, we introduce a first strictly efficient scheme relying only on a simple controlled rotation, which can be realized by an atom-light interaction of the Jaynes-Cummings type or purely optically by the Kerr effect,  for the creation of the resource. On the measurement side, we require beamsplitters, phase-space displacers, and photon counting measurements. While this scheme is not fleshed out in all detail of its concrete physical implementation, it should be clear that quantum optical implementations of such ideas are conceivable.

The present article is organized as follows: First, we discuss what operations are possible with continuous-variable quantum systems and introduce a framework based on matrix product states (MPS) to describe MBQC in a general setting. After a discussion of the properties which a CV-MBQC scheme needs to posses in order to be called theoretically efficient, we show the problems of achieving  those requirements with Gaussian measurements on non-Gaussian resource states. Last, we provide an example for a feasible scheme and discuss in detail how efficient MBQC can be performed in this situation. 

\section{Feasible primitives for CV quantum computing}
When performing tasks of quantum information with continuous-variable quantum optics different classes of operations are considered to be of different difficulty. This is also true for measurements where the achievable efficiency greatly differs between the various methods. 

\subsection{Gaussian operations}
\label{sus:gaussop}
Before continuing with the discussion, we remind the reader of some basic properties of Gaussian states and operations while also taking the opportunity to set the notation. We consider a single light mode. The energy eigenvectors of the unit oscillator are denoted by $|n\rangle$ with $n=0,1,\ldots$. The annihilation operator $\hat{a}$ acts on them according to $\hat{a}|n\rangle=\sqrt{n}|n-1\rangle$. The commutator relation with its adjoint, the creation operator, is $[\hat{a},\hat{a}^\dag]=1$, setting $\hbar=1$. The eigenvectors of the photon-number operator $\hat{n}=\hat{a}^\dag\hat{a}$ with $\hat{n}|k\rangle=k|k\rangle$ are called the number states or Fock states. One can now define the canonical operators or quadratures as
\begin{eqnarray}
	\hat{q}&=&\frac{1}{\sqrt{2}}(\hat{a}+\hat{a}^\dag), \\
	\hat{p}&=&-\frac{i}{\sqrt{2}} (\hat{a}-\hat{a}^\dag), 
\end{eqnarray}
and for $\theta\in [0,2\pi]$ the family of rotated quadrature operators 
\begin{equation}	
\label{eq:rotq}
	\hat{q}_\theta=\hat{q}\cos(\theta)+\hat{p}\sin(\theta) .
\end{equation}
The latter family of observables is the one that captures homodyne detection \cite{schleich}. Gaussian unitaries are the ones which can be written as $\hat{U}=e^{i\hat{H}(\hat{q},\hat{p})}$ where $\hat{H}$ contains no terms in higher than quadratic order in $\hat{q}$ and $\hat{p}$ (or, equivalently, in $\hat{a}$ and $\hat{a}^\dag$). There are three classes of Gaussian single-mode unitary operators into which all Gaussian unitary gates can be decomposed. The first ones are corresponding to the application of the displacement operator $\hat{D}(\alpha)=\exp(\alpha\hat{a}^\dag-\alpha^*\hat{a})$ 
for $\alpha\in\cc$. Such a  transformation is reflected in the Heisenberg picture by a map of the form
\begin{eqnarray}
	\hat{q}&\mapsto&\hat{q}+\sqrt{2}\,{\rm Re}\,\alpha ,\\
	\hat{p}&\mapsto&\hat{p}+\sqrt{2}\,{\rm Im}\,\alpha.
\end{eqnarray}	
Optically, 
such a transformation can be realized by mixing the mode with a second mode, which is in a strong coherent state, on a beam splitter in the limit of vanishing reflectivity. In the second one are the transformations generated by
the clockwise rotation operator $R(\theta)=\exp(-i\theta\hat{n})$, which can be realized by a phase shifter, acting on the canonical operators as $\hat{q}\mapsto\hat{q}_\theta$, $\hat{p}\mapsto\hat{q}_{\theta+\pi/2}$, while in the last one are those
generated by the squeezing operator 
\begin{equation}
	\hat{S}(\xi)=\exp\left[\frac{r}{2}(\hat{a}^2-\hat{a}^{\dag 2})\right], 
\end{equation}
acting as $\hat{q}\mapsto e^r \hat q$, $\hat{p}\mapsto e^{-r}\hat{p}$. The single-mode Gaussian unitary operations form a (non-compact) group which we denote by $\mathbb{U}_{\rm G}$. To complete the set of operators, which are necessary to implement arbitrary \textit{multi-mode} Gaussian operations, we introduce the (absorption-free) beam splitter acting on two modes $1$ and $2$ by 
\begin{equation}
	\hat{B}=\exp\left[\frac{\theta}{2}(\hat{a}_1^\dag\hat{a}_2-\hat{a}_1\hat{a}_2^\dag)\right]
\end{equation}
where $t=\cos(\theta/2)$ and $r=\sin(\theta/2)$ are the transmission and reflection coefficient, respectively. 

The Gaussian operations can be divided into two classes: Phase shifters and beam splitters do not change the total number of photons and are called passive operations. To implement a squeezer or a displacer, on the other hand, one does need additional photon sources which makes them more difficult to realize. 

\subsection{Other Hamiltonian building blocks}
There are also non-Gaussian operations which are within the realm of present experimental techniques: The cross Kerr effect gives rise to a non-linear coupling between two modes; its Hamiltonian reads
\begin{equation}
\label{eq:crossKerr}
\hat{H}=\chi\hat{n}_1\otimes\hat{n}_2.
\end{equation} 
It can be realized by transmitting two modes together through an optically non-linear medium. As non-linear optical effects are weak and absorption needs to be kept low, the achievable effective coupling strength is quite small, i.e., $t\chi\ll 1$, where $t$ is the interaction time. For the most relevant situation, where one of the modes carries a state with many photons while the other one is in the single photon regime, the Hamiltonian (\ref{eq:crossKerr}) has already been experimentally implemented to perform quantum non-demolition measurement (QND) \cite{crosskerr}. The related Hamiltonian
\begin{equation}
\label{eq:atomfield}
\hat{H}=\chi |1\rangle\langle 1|\otimes \hat{n}
\end{equation}
for a positive $\chi$
describes the coupling of a two-level atom to a light mode in the dispersive limit of the Jaynes-Cummings model. In this situation, the effective coupling strength can be increased by placing the mode and the atom inside an optical cavity. A particularly strong interaction can be achieved with a technique known as electromagnetically induced transparency (EIT) which is routinely used to exchange quantum information between between light modes and atomic vapors and for which also experiments with single atoms exist \cite{singleatom1,singleatom2}.

\subsection{Measurements}

We now turn to the kind of measurements that we will consider feasible for the purposes of this work.
The most important Gaussian measurement scheme is homodyne dectection which correponds to the observable $\hat{q}_\theta$ in Eq.~(\ref{eq:rotq}). It is realized by combining the mode with a strong laser, called the local oscillator, in an interferometer, measuring the intensities on both output ports, and subtracting the results. Another
important type of Gaussian measurement is eight-port homodyning corresponding to a direct
measurement of the Q-function, i.e.,\,$Q_\rho(\alpha)=\langle\alpha|\rho|\alpha\rangle$ for $\alpha\in\cc$, where 
\begin{equation}
\label{eq:defalpha}
	|\alpha\rangle=e^{-|\alpha|^2/2}\sum_n\frac{\alpha^n}{\sqrt{n!}}|n\rangle
\end{equation}	
are the state vectors of the non-orthogonal and overcomplete coherent states \cite{schleich}.

Non-Gaussian measurements are in many instances 
more difficult to realize than Gaussian ones, and usually with significantly lower detection efficiencies. The most
feasible instance of a non-Gaussian measurement reflects a 
single-photon detector which can only distinguish between the absence of photons and the presence of one or more photons. The corresponding POVM-elements are $|0\rangle\langle 0|$ and $\id-|0\rangle\langle 0|$. A photon-number resolving detector is more challenging to implement but with time-multiplexing \cite{multiplexing} or superconducting nano-wires \cite{nanowires}, it is possible to perform photon counting for the first few number states with reasonable efficiency. 

\section{Framework for CV-MBQC}
\label{sec:framework}
In this section, we introduce a general framework to describe measurement-based quantum computing.
A MBQC scheme consists two ingredients: First, a resource state and second a set of possible, or allowed local measurements. We start with a description of quantum wires which are used for single-qudit processing \cite{webs} and discuss their coupling to fully universal resources afterwards. 

 \subsection{Matrix product states}
\label{sus:mps}
The formalism of matrix product states, which was originally introduced to describe a certain class of entangled one-dimensional many-body states \cite{mpstheory1,mpstheory2} is extremely useful to capture the essentials of quantum computing in the measurement based model \cite{pra}. We start by describing one-dimensional continuous-variable quantum wires which can be used to carry a single qubit of quantum information. The techniques we develop are independent of the actual physical implementation but we will often describe them with terms of quantum optics and have such a system in mind. 

We consider a chain of $L$ quantum systems with dimension $d_{\rm p}$ which are called lattice sites. We also allow for the situation $d_{\rm p}=\infty$ which describes CV light modes. We say that a state is physical if it has finite mean energy, i.e., for a single mode: $\Tr\left(\hat{n}\rho\right)\le\infty$. We choose a basis (or a countably infinite Hilbert-space basis) $\{|i\rangle:i=1,\ldots,d_p\}$ of $\cc^{d_p}$, which we call the computational basis, associate to any basis element a $D$-dimensional matrix $A[i]\in\mathcal{M}_D(\cc)$, and write a translationally invariant matrix product state (MPS) as 
\begin{equation}
|\Psi\rangle=\sum_{i_L,\ldots,i_2,i_1=1}^{d_p}\langle L|A[i_L]\ldots A[i_2]A[i_1]|R\rangle|i_L,\ldots,i_2,i_1\rangle
\label{eq:mpsdef}
\end{equation}
where $|L\rangle\,,|R\rangle\in\cc^D$. This vector space is the one where the quantum computation will take place and is called correlation space. The matrices must fulfill the completeness relation 
\begin{equation}
\sum_{i=1}^{d_p}A[i]^\dag A[i]=\id,
\end{equation}
for rendering deterministic computation feasible.

An important fact is that all physical states can be approximated arbitrary well by a finite-dimensional MPS. As any finite-dimensional state can be written as a MPS \cite{mpstheory2}, it only remains to show that one can truncate states with finite energy. We write the mean total photon number of a state $\rho$ with $k$ modes as
\begin{align}
N_{\rm mean}=&\sum_{n_1,\ldots,n_k=0}^\infty(n_1+\ldots+n_k)\rho_{n_1,\ldots,n_k;n_1,\ldots,n_k}\nonumber\\
\ge&k(n_{\rm max}+1)\sum_{n_1,\ldots,n_k=n_{\rm max}+1}^\infty \rho_{n_1,\ldots,n_k;n_1,\ldots,n_k},\label{eq:totalmeanphoton}
\end{align}
where the matrix elements of $\rho$ are the ones of the Fock basis. 

Let $\rho_{\rm trunk}$ be the non-normalized state obtained from $\rho$ by setting $\rho_{n_1,\ldots,n_k;n_1',\ldots,n_k'}=0$ if one of the indices is larger than $n_{\rm max}$. This state fulfills
\begin{equation}
\label{eq:meanphotontemp}
\Tr(\rho-\rho_{\rm trunc})\le\frac{N_{\rm mean}}{k(n_{\rm max}+1)}.
\end{equation}
Denoting the trace norm, which is the relevant norm for the distinguishability of quantum states, by $\|\cdot\|_1$, we get from Ref.~\cite{cvcs}
\begin{equation}
\label{eq:meanphoton}\|\rho_{\rm trunc}-\rho\|_1\le3\left({\frac{N_{\rm mean}}{k(n_{\rm max}+1)}}\right)^{1/2}.
\end{equation}

 If we assume the mean photon number (the energy) \textit{per mode} to be constant, the truncation error is independent of the system size. This means, for a given required accuracy, we can choose a cut-off number $n_{\rm max}$ independently of the length of the intended computation.

\subsection{MPS as quantum wires}
We now show how single-qudit MBQC can be performed with such a matrix product state, following and extending the line of reasoning used in Ref.~\cite{pra}. Let us assume, the first mode is measured in the computational basis with result $k$. Then, the remaining system is described by the state vector
\begin{equation}
|\Psi\rangle\propto\sum_{i_L,\ldots,i_2=1}^{d_p}\langle L|A[i_L]\ldots A[i_2]A[k]|R\rangle|i_L,\ldots,i_2\rangle.
\label{eq:mpsdef2}
\end{equation}
This can be interpreted as the action of the matrix $A[k]$ on the right boundary vector $|R\rangle$. To interpret this as the action of a quantum gate it is necessary that $A[k]$ is proportional to a unitary matrix, i.e., $A[k]^\dag A[k]=p(k)\id$ where $p(k)$ is the probability with which the measurement result $k$ is obtained. When the measurement basis is continuous, we denote by $p$ the probability density while $P$ denote the corresponding probability measure. Because we will also discuss measurements where the corresponding matrices are not proportional to unitary ones, we note that the probability in this more general situation is $p(k)=\|A[k]|\psi\rangle\|^2$, where $|\psi\rangle$ is the state vector of the correlation system. If the measurement is not performed in the computational basis but in another one and the result corresponds to a projection to the state vector $|x\rangle$, the matrix applied on $|R\rangle$ reads 
\begin{equation}
	A_{\cal B}[x]=\sum_{i=1}^{d_p}\langle x|i\rangle A[i]\,.
\label{eq:skewmeasurement}
\end{equation}
Note that the basis $\mathcal{B}=\{|x\rangle\}$ may be continuous and/or overcomplete and that measurements in such bases naturally occur in CV-quantum optics. We also allow this basis to consist of improper eigenstates reflecting an idealized homodyning detection. We can now already note an almost trivial but very important necessary requirement for a single-qudit MBQC scheme: If there exists no allowed measurement basis such that $A[x]$ is proportional to a unitary matrix for almost all $x$, it is not even possible to transport a single $D$-dimensional qudit.  An equivalent formulation is that $p(x)$ must not depend on the state vector $|\psi\rangle$ of the correlation space. If this was the case, a measurement would yield information about $|\psi\rangle$, which would clearly destroy coherence.

\subsection{Sequential preparation}
\begin{figure}
\begin{center}
\includegraphics{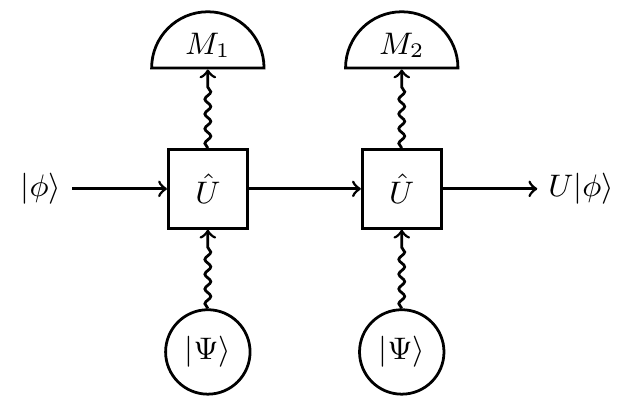}
\end{center}
\caption{\label{fig:seqprep} Interpretation of single-qudit MBQC as sequential interaction with an ancilla. The physical sites are initially in the state vector $|\Psi\rangle$ and interact with the correlation system through $\hat{U}$. Depending on the measurement results in $M_1$ and $M_2$, the unitary gate $U$ is applied on $|\phi\rangle$.}
\end{figure}

A possible way of preparing a matrix product state with bond dimension $D$ is a sequential interaction with a $D$-dimensional auxiliary system \cite{mpstheory2,sequentialpreparation}. This picture proves to be extraordinarily 
useful when discussing quantum wires. Let the interaction between this auxiliary system and a local physical system be described by
\begin{equation}
\hat{U}:\cc^D\otimes\cc^{d_p}\rightarrow\cc^D\otimes\cc^{d_p}
\label{eq:interactionunitary}
\end{equation}
and assume the physical system to be initialized in the state vector $|\Psi\rangle$. Then, the matrix elements of the MPS matrices read
\begin{equation}
\label{eq:Aseqinteraction}
A[i]_{j,k}=\left(\langle k|\otimes\langle i|\right)\hat{U}\left(|j\rangle\otimes|\Psi\rangle\right).
\end{equation}
In this picture, which is sketched in Figure \ref{fig:seqprep}, the correlation system is identified with the auxiliary system, and a single qudit gate is performed by first letting the auxiliary system interact with a mode in a known state and then measuring the mode. It is important to note that this picture works regardless of the actual way the resource state is prepared. For this reason, we will use the picture of sequential preparation to be more intuitive while maintaining full generality. For example, resource states can also arise as ground states of local Hamiltonians or be prepared by the action of translationally invariant quantum cellular automata \cite{groundstates1,groundstates2}.

\subsection{Encoding of quantum information in correlation spaces}
\label{sus:encoding}
If the goal is not to process a $D$-dimensional qudit but only a $d$-dimensional one with $d<D$, the above requirement is too strong. The $d$-dimensional space is called the computational subspace. We introduce the encoding
\begin{equation}
\label{eq:encoding}
V:\cc^d\rightarrow\cc^D
\end{equation}
with $V^\dag V=\id_d$ and the encoded matrices
\begin{equation}
B[i]=V^\dag A[i]V\,,\quad B\in\mathcal{M}_d(\cc)\,.
\label{eq:encA}
\end{equation}
If they are proportional to unitary matrices for all $i$, it is possible to process a single $d$ dimensional qudit. One can observe that 
\begin{multline}
\label{eq:encB}
\langle \phi|V^\dagger A_{b_L}[i_L]\ldots A_{b_1}[i_1] V|\psi\rangle\\
=\langle\phi| B_{b_L}[i_L]\ldots B_{b_1}[i_1]|\psi\rangle
\end{multline}
where $|\phi\rangle,|\psi\rangle \in\cc^d$. The subscripts $b_i$ denote the chosen measurement basis. This means, for all measurements, the state with the matrices $B$ behaves exactly like the one with the matrices $A$. Thus, we can just proceed as if we had a state with a $d$-dimensional correlation space from the very beginning. 

This situation is not the most general one. The isometry (\ref{eq:encoding}) can depend on the measurement basis and the outcome of the previous steps. In this case, the computational subspace is not fixed but changes during the computation. It is reasonable to demand that at least the initial and the final encoding coincide so that it is possible to work with a fixed in- and out-coupling, provided by $V$. In this case Eq.~(\ref{eq:encA}) becomes 
\begin{multline}
\label{eq:enB}
\langle \phi|V^\dagger A_{b_L}[i_L]V_{b_{L-1}}[i_{L-1}]\ldots V_{b_1}[i_1]V^\dag_{b_1}[i_1]A_{b_1}[i_1]V|\psi\rangle\\
=\langle\phi| B_{b_L}[i_L]\ldots B_{b_1}[i_1]|\psi\rangle
\end{multline}
where $ B[i_k]=V^\dag_{b_k}[i_k]A[i_k]V_{b_{k-1}}[i_{k-1}]$. When several quantum wires are used to perform a quantum computation, it is also necessary to demand the changing encoding to return to the initial encoding given by $V$, whenever a coupling occurs. If this was not the case, the couplings would have to depend on the history of measurements and their results in both of the wires. This would clearly be against the spirit of a measurement-based protocol, where all the adaptation lies in the choice of measurement bases only. 

The infinite-dimensional matrix product states used in this work bear some resemblance to the continuous matrix product states which where introduced in Refs.\ \cite{cMPS1,cMPS2} to describe one-dimensional bosonic quantum fields. In contrast to the discrete structure of bosonic modes described here, they are formulated without an underlying lattice. However, such cMPS can be obtained by a suitable continuum limit from infinite-dimensional matrix product states. In this limit, only two independent MPS matrices remain while the rest can be calculated from them which is an important difference to the present situation.

\subsection{Coupling of wires}
\label{sus:couplewires}

\begin{figure}
\begin{center}
\includegraphics{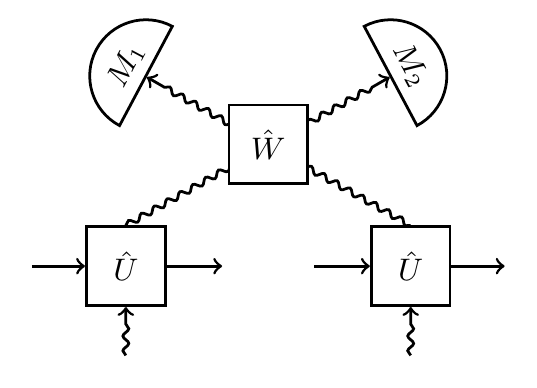}
\end{center}
\caption{\label{fig:coupling} Coupling of two quantum wires. The interaction of two physical systems through $\hat{W}$ and the subsequent measurements ($M_1$ and $M_2$) induces a entangling gate on the two correlation systems.} 
\end{figure}

To go beyond single-qudit processing and achieve quantum computational universality, wires have to be coupled. Physically, this can be performed in different ways. One possibility is to let the auxiliary systems of the sequential preparation scheme interact. However, this might be very challenging to do in reality, both when working with atoms, due to the difficulty of controlling them, and with light, due to weak non-linear interaction. For this reason, joint measurements are preferable to couple two wires which is also shown in Figure \ref{fig:coupling}. In an optical situation, this could mean to combine two modes belonging to two different wires on a beam splitter and measuring both output ports.  This is a broadening of the usual definition of MBQC where only local measurements are performed while we now also allow for two-local ones. On the other hand, there also exists a closely related third way which is also based on the beam-splitter interaction but strictly stays within the measurement-based paradigm. The idea is to perform the coupling independently of the executed algorithm according to some fixed scheme. This step then belongs to the preparation of the resource while the subsequently performed measurements are purely local.

Let $\hat{W}$ be the coupling between the two modes, $\{|x_1\rangle\}$ and $\{|x_2\rangle\}$ the measurement bases after interaction. Then, the applied operation reads 

\begin{equation}
\label{eq:twosite}
A[x_1,x_2]=\sum_{i,j}\langle x_1,x_2|\hat{W}|i,j\rangle A[i]\otimes A[j].
\end{equation}

\section{Requirements for efficient MBQC schemes}
\label{sec:requirements}

In this Section, we give three condition a MBQC-scheme has to fulfill to be called efficient and comment on their importance. To facilitate the derivation, we restrict ourselves to the, arguably most important, case of $D=2$ while stressing that the requirements for universal computing on qudits with larger dimension are completely analogous. When discussing requirements, one has to distinguish between two different notions: One the one hand, there are \textit{theoretical} requirements, of the kind that, if they are met, efficient scalable 
quantum computing is possible in the absence of noise. Going away from this idealized situation, on the other hand,
two new features arise. Firstly, in practice, any quantum operation is affected by noise, which error-correction may
allow to counter to some extent. Secondly, any actual quantum computation is, needless to say, 
finite. Thus, the practical requirements that might be sufficient to sustain a computation
may differ from the theoretical ones. However, as we are primarily interested in 
scalable quantum computing and for reasons of clarity we will focus on the 
theoretical requirements and comment on other practical implications 
whenever it is in order.

\subsection{Conditions}

We start with the theoretical requirements for quantum wires (which process single qubits) and state the first necessary condition. 
\begin{requirement}[Transport]
\label{req:requirement}
Let $|\psi_n \rangle$ be the state vector of the correlation system after $n$ steps in the quantum wire when the initial state vector was $|\psi\rangle$. We say that transport of a single qubit is possible if there exists some $\eps>0$ such that for all $n$ 
\begin{equation}
\label{eq:req}
	F_{\rm min}:=\max_{U\in\mathbb{SU}_2}\min_{|\psi\rangle}
	|\langle\psi|U|\psi_n \rangle|^2\ge\frac{1}{2}+\eps.
\end{equation}
\end{requirement}
Here, $U$ is the best attempt at undoing the action of the quantum wire on $|\psi\rangle$. This requirement ensures that computations are not limited in length. If encodings of single logical qubits in multiple wires are used, this requirement can be weakened to ${1/\eps={\rm poly}(n)}$. 

For universal single-qubit quantum computing to be possible, it is not enough to merely transport quantum information. It is also necessary that any one-qubit unitary $U\in\mathbb{SU}_2$ can be efficiently approximated. Because it depends on the measurement outcome which operation is applied, we cannot hope for this to be possible deterministically. Instead, the measurements induce a random walk on $\mathbb{SU}_2$ which can be controlled by choosing the measurement basis in the next step depending on the gate implemented so far. For efficient quantum computing, the average number of steps must not increase too fast with the desired accuracy. We set 
\begin{equation}
\bar{A}_\mathcal{B}[k]:= {p(k)}^{-1/2}A_\mathcal{B}[k]
\end{equation}
where $\mathcal{B}$ denotes the measurement basis and define 
\begin{equation}	
	\bar{A}_\mathcal{B}[\mathbf{k}]=\prod_{i=1}^n\bar{A}_{\mathcal{B}_i}[k_i].
\end{equation}	
\begin{requirement}[Single qubit universality]
\label{req:approximation}
Let $U\in\mathbb{SU}_2$ and $\eps>0$. There exits a sequence (possibly adaptive) of bases $\mathcal{B}_i$ such that the expectation value for an approximation $\tilde{U}=\bar{A}_\mathcal{B}[\mathbf{k}]$ fulfilling 
\begin{equation}
\|\tilde{U}-U\|\le\eps
\end{equation}
after at most $n$ measurements in the wire satisfies
\begin{equation}
\EE\left(n\right)=O\left({\polylog}\left(\frac{1}{\eps}\right)\right).
\label{eq:approximation}
\end{equation}
\end{requirement}
Here, $\|\cdot\|$ denotes the operator norm, which is the meaningful figure of merit in this case. 
Of course, single qubit operations are not sufficient for universality but an entangling operation is also needed:

\begin{requirement}[Coupling]
\label{req:coupling}
Denote with
\begin{equation}
	C_Z={\rm diag}(1,1,1,-1)
\end{equation}
the controlled-$Z$ gate. The expectation value for an approximation $\tilde{C}_Z$ fulfilling 
\begin{equation}
	\|\tilde{C}_Z-C_z\|\le\eps
\end{equation}
 after at most $n$ measurements on both of the quantum wires involved, where both single-site and two-site measurements are allowed, must satisfy
\begin{equation}
\label{eq:coupling}
\EE\left(n\right)=O\left({\polylog}\left(\frac{1}{\eps}\right)\right).
\end{equation}
\end{requirement}
Requirements \ref{req:approximation} and \ref{req:coupling} guarantee that every quantum circuit can be simulated in correlation space with a polylogarithmic overhead. This is the commonly used requirement in quantum computing. When one is content with a polynomial overhead, (\ref{eq:approximation}) and (\ref{eq:coupling}) can both be relaxed to
\begin{equation}
\label{eq:coupling2}
\EE(n)=O\left(\poly\left(\frac{1}{\eps}\right)\right)
\end{equation}
for some constant $c>0$. A computational model fulfilling only this weaker condition is still of interest when the ultimate aim is to perform a quantum algorithm which provides exponential speed-up over the corresponding classical one (like Shor's algorithm). However, whenever the speed-up is only polynomial (like in Grover's search algorithm), the stronger polylog scaling is clearly necessary for the quantum algorithm to have any advantage. When one aims at the implementation of constant size quantum circuits only, Requirements 2 and 3 are not necessary as one does not need to implement the quantum gates to arbitrary accuracy but only to some $\eps>0$ which is determined by the total length of the computation.
 
The last feature to demand from a MBQC model is the ability to initialize the correlation system into some known state and to perform a measurement of the correlation space in the computational basis.

\begin{requirement}[Initialization and read-out]
\label{req:iniandreadout}
For efficient initialization of the correlation system to be possible, there must exist some measurement sequence such that, independently from its initial state vector $|\psi\rangle$, 
\begin{equation}
	\||\psi_n \rangle-|0\rangle\|\le\eps. 
\end{equation}
Here $|\psi_n \rangle$ is the state vector after $n$ steps in the wire and it must be true that 
$\EE(n)=O\left(\polylog(1/\eps\right))$.

For readout, it is necessary that after $n$ steps with 
\begin{equation}
\EE\left(n\right)=O\left(\polylog\left(\frac{1}{\eps}\right)\right)
\end{equation}
the combined normalized action 
\begin{equation}
	\bar{A}_\mathcal{B}[\mathbf{k}] =\prod_{j=n}^1\bar{A}_{\mathcal{B}_j}[\mathbf{k_j}] 
\end{equation}
 fulfills either $\|\bar{A}_\mathcal{B}[\mathbf{k}]-|0\rangle\langle 0|\|\le\eps$ or $\|\bar{A}_\mathcal{B}[\mathbf{k}]-|1\rangle\langle 1|\|\le\eps$. 
\end{requirement}
The first requirement just means that one can approach one of the computational basis states fast enough, while the second one ensures an efficient implementation of an approximative measurement in the computational basis. While measurement-based 
computational schemes that do not respect these requirements may be conceivable in principle, within the framework 
presented here they are very natural indeed and necessary for universal quantum computing.

\subsection{Consequences}
In the previous section we have discussed the requirements a MBQC scheme has to fulfill in order to be called efficient. We now show some consequences of these requirements which will help us both to find classes of schemes which cannot be efficient and such that are. This seems an important enterprise in order to arrive at
``no-go results'', so to identify serious limitations that have to be circumvented in one way or the other.
For many important situations, $F_{\rm min}$ in Requirement \ref{req:requirement} vanishes exponentially in $n$ as the following observation shows, which by virtue of the above statement implies that efficient quantum computing is not possible.

\begin{observation}[Impossibility of transport]
\label{obs:transport}
Given a basis $\mathcal{B}$, let $\mathcal{C}\subset\mathcal{B}$, with $P(\mathcal{C})>0$. If for all $x\in \mathcal{C}$, $A[x]\not\propto U$ with $U\in\mathbb{SU}_2$, transport is impossible when measuring in this basis.
\end{observation} 
Proof: We calculate the fidelity
\begin{equation}
\label{eq:fmin}
f_{\rm min}(x):=\max_{U[x]\in\mathbb{SU}_2}\min_{|\psi\rangle}\frac{|\langle\psi|U[x]A[x]|\psi\rangle|^2}{p(x)}.
\end{equation}
Because $P(\mathcal{C})>0$ and $f_{\rm min}(x)<1$ for all $x\in\mathcal{C}$, there exists a non-empty compact interval $\mathcal{I}\subset\mathcal{C}$ such that there is some $\delta<1$ with $f_{\rm min}(x)\le\delta$ for all $x\in\mathcal{I}$. Using an argument from Ref.\ \cite{nogo}, one can show that this implies the exponential decay of $F_{\rm min}$, showing that Requirement \ref{req:requirement} is not satisfied. Thus, for transport to be possible, there must exist a basis such that almost all measurement outcomes correspond to some $A[x]$ with $A[x]^\dag A[x]\propto\id$. For transport over a fixed length, Observation \ref{obs:transport} does not apply because even if the fidelity decays exponentially fast, it might be still large enough for the task under question. 

We show a simple sufficient condition, which will be used later when discussing examples for which Requirement \ref{req:approximation} is true:
\begin{observation}[Condition for an efficient random walk]
\label{obs:randwalkcond}
Assume that for any target gate $U\in\mathbb{SU}_2$ there exists some basis $\mathcal{B}$ and some $\mathcal{C}\subset\mathcal{B}$ for any $P(\mathcal{C})\ge p(\eps)$ with $1/p(\eps)=O\left(\polylog(1/\eps)\right)$ where $\eps$ is the desired accuracy. When 
\begin{equation}
	\|\bar{A}_\mathcal{B}[k]-U\|\le\eps
\end{equation}
for all $k\in\mathcal{C}$, Requirement \ref{req:approximation} is fulfilled.
\end{observation}
Proof: The expected number of tries can be bounded by
\begin{align}
\EE(n_\eps)\le&\sum_{k=1}^\infty kp(\eps)(1-p(\eps))^{k-1}=\frac{1}{p(\eps)}\nonumber\\
=&O\left(\polylog\left(\frac{1}{\eps}\right)\right).
\label{eq:numtries}
\end{align}
The most important important situation covered by Observation \ref{obs:randwalkcond} is the one where the measurements have discrete results and for every $U\in\mathbb{SU}_2$ a basis exists which contains at least a single element which fulfills the requirements of Observation \ref{obs:randwalkcond}.
Sufficient conditions for initialization and read-out of the correlation system to be possible are provided by the following observation:

\begin{observation}[Sufficient conditions for Requirement \ref{req:iniandreadout}]
\label{obs:iniandreadout}
Let there exist some $\delta>1$ and a basis $\mathcal{B}$ containing some subset $\mathcal{C}\subset\mathcal{B}$ of ``non-unitary measurements''. Assume that 
\begin{equation}
A[x]\propto U[x]\left[\begin{array}{cc} s_1(x) & 0 \\ 0 & s_2(x) \end{array}\right]V^\dag
\end{equation}
for all $x\in\mathcal{C}$, where $U[x],V\in \mathbb{SU}_2$ and $s_1(x)$, $s_2(x)$ are the (not necessarily ordered) singular values of $A[x]$ which fulfill either $s_1/s_2>\delta$ or $s_2/s_1>\delta$. In this case, efficient read-out and initialization of the correlation system is possible.
\end{observation} 
Proof: The situation is most transparent when $s_2(x)=0$. In this case, a measurement result $x$ corresponds to a projection of the correlation space state to the state vector $V|0\rangle$. In the more general case the measurement is not projective but merely weak. By performing the unitary operation $VU^\dag[x]$, which is efficiently possible due to Requirement \ref{req:approximation}, and repeating the measurement, a projective measurement in the basis $\{V|0\rangle,V|1\rangle\}$ can be approximated. This approximation is efficient, due to the existence of a finite gap between $s_1$ and $s_2$ occurring with finite probability. The independence of $V$ of $x$ is crucial: If this is not the case, the basis in which the correlation space measurement will occur cannot be fixed. This will lead to a destruction of quantum information resulting in a failure of the MBQC scheme. Initialization of the correlation system in the state vector $|0\rangle$ can be performed in the very same manner. One just performs the measurement procedure outlined above after which the correlation system is in $U[x]|0\rangle$ or $U[x]|1\rangle$, depending on the outcome. Applying now the gates $U[x]^\dag$ or $XU[x]^\dag$, respectively, achieves the required initialization. 

\section{Limitations}

\label{sec:limitations}
When devising a MBQC scheme, there are several fundamental limitations concerning the use of Gaussian states and Gaussian measurements. The first important fact is that all protocols involving only Gaussian measurements on Gaussian states can be simulated efficiently on a classical computer, ruling out the possibility for any quantum speed-up \cite{efficientsimulation}. However, for an \textit{ideal} Gaussian cluster state, universality can be achieved by adding a single non-Gaussian measurement to the toolbox \cite{gaussiancluster1}. On the other hand, any \textit{physical} Gaussian quantum wire, including an one-dimensional Gaussian cluster state cannot fulfill Requirement \ref{req:requirement} even when allowing for non-Gaussian measurements \cite{nogo}.
\subsection{Controlled Gaussian operations}
\label{sus:controlledG}

We now discuss further limitations and consider the situation where the interaction (\ref{eq:interactionunitary}) can be written as
\begin{equation}
\hat{U}=|0\rangle\langle 0|\otimes\hat{G}_0+|1\rangle\langle 1|\otimes\hat{G}_1,
\label{eq:controlledGaussian}
\end{equation}
up to local unitaries on the correlation system, where $\hat{G}_0,\hat{G}_1\in\mathbb{U}_G$. This class contains states which allow for transport of a qubit in a wire by Gaussian measurements only. An example is given by 
\begin{equation}
\label{eq:controlledrot}
\hat{U}=|0\rangle\langle 0|\otimes\exp(-i\theta\hat{n})+|1\rangle\langle 1|\otimes\exp(i\theta\hat{n})
\end{equation}
where $\theta>0$ is a parameter characterizing the interaction strength. Taking a coherent state vector $|\alpha\rangle$ with $\alpha\in\rr$, $\alpha>0$ as input, $q_0=\sqrt{2}\alpha\cos(\theta)$, $p_0=\sqrt{2}\alpha\sin(\theta)$, a $q$-quadrature measurement with result $x$ applies in correlation space
\begin{equation}
\label{eq:Aq}
\bar{A}[x]={\rm diag}\,(e^{ip_0(x-\frac{q_0}{2})},e^{-ip_0(x-\frac{q_0}{2})})
\end{equation}
with probability $p(x)=\exp(-(x-q_0)^2)/\sqrt{\pi}$. Thus, Requirement \ref{req:requirement} is fulfilled. However, efficient single-qubit gates are impossible due to the lack of control. The only possible measurement (up to squeezing) for which $\bar{A}[x]$ is unitary are phase-space displacements followed by $q$-quadrature homodyne detection as for all others, the projectors do not fulfill 
\begin{equation}
	|\langle\psi(x)|e^{i\theta}\alpha\rangle|=|\langle\psi(x)|e^{-i\theta}\alpha\rangle|
\end{equation}
which implies that the probabilities of the measurements depend on the state of the correlation system and, therefore, the applied matrix is not unitary. Writing the displacement as $\Delta=(\Delta_q+i \Delta_p)/\sqrt{2}$, we get
\begin{equation}
\label{eq:ttt3}
\bar{A}_\Delta[x]={\rm diag}\,(e^{i(p_0+\Delta_p)(x-\frac{q_0+\Delta_q}{2})},e^{-i(p_0-\Delta_p)(x-\frac{q_0+\Delta_q}{2})})
\end{equation}
with probability $p_\Delta(x)=\exp(-(x-q_0-\Delta_q)^2)/\sqrt{\pi}$. Up to a redefinition $q_0\mapsto q_0+\Delta_q$ and some global phase, (\ref{eq:ttt3}) is constant in $\Delta$, i.e., there is no way of controlling the random walk, making it impossible to meet Requirement \ref{req:approximation}. The same argument holds for any controlled Gaussian operation and for any Gaussian input state. 

When allowing for non-Gaussian input states, some amount of control is indeed possible, as the following example shows:  Consider the interaction of Eq.~(\ref{eq:controlledrot}) together with a superposition of two different photon numbers 
as an input, $|\psi\rangle=(|0\rangle+|2\rangle)/\sqrt{2}$. Choosing $\theta=\pi/4$, the post-interaction state reads
\begin{equation}
a|0\rangle\frac{1}{\sqrt{2}}\left(|0\rangle -i|2\rangle\right)+b|1\rangle\frac{1}{\sqrt{2}}\left(|0\rangle +i|2\rangle\right)
\end{equation}
and the normalized matrices $\bar{A}_q$ and $\bar{A}_p$ for the $q$- and $p$-quadrature measurements are
\begin{align}
\bar{A}_q[x]=\,&{\rm diag}\,(e^{-i\phi(x)},e^{i\phi(x)}),\\
\bar{A}_p[x]=\,&{\rm diag}\,(e^{i\phi(x)},e^{-i\phi(x)}),
\end{align}
with 
\begin{equation}
	\phi(x)={\rm arctan}\,\frac{2x^2-1}{\sqrt{2}}
\end{equation}
 and 
 \begin{equation}
 p(x)=(4x^4-4x^2+3)\exp(-x^2)/(4\sqrt{\pi}). 
 \end{equation}
 Thus, there exist two Gaussian measurements which allow to control the random walk but even this control is not enough to meet Requirement \ref{req:approximation}. Because this is a general feature of any MBQC-scheme relying solely on Gaussian measurements we will discuss it in detail.

\subsection{General resource states with Gaussian measurements}
\label{eq:susgeneralresources}
We consider an MPS with $D=2$ with the continuous family of matrices $A_q[x]$ with $x\in\rr$. As the projectors describing eight-port homodyning are not orthogonal, they will turn the state of the remaining system into a mixed one when measuring a mode, and we do not need to consider this  measurement. Because  all relevant Gaussian measurements can be expressed as the application of squeezing, rotating, and displacing, followed by a $q$-quadrature measurement, we can write
\begin{equation}
\label{eq:allgauss}
A_{\theta,\Delta_q,\Delta_p,\lambda}[x]=\cos(\theta)A_q[\lambda x+\Delta_q]+\sin(\theta)A_p[x/\lambda+\Delta_p]
\end{equation}
where 
\begin{equation}
	A_p[q]=\frac{1}{\sqrt{2\pi}}
	\int{\rm d}x\,A_q[x]e^{iqx},
\end{equation}
and where we have omitted global phase factors. If single-qubit transport is possible in this wire, we can, without loss of generality, assume that $A_q[x]:=A_{0,0,0,0}[x]$ is proportional to a unitary matrix for all relevant $x$.  As both squeezing by a finite $\lambda$ and shifting in phase space does merely result in a redefinition of $x$ in Eq.~(\ref{eq:allgauss}), we can restrict ourselves to the discussion of
\begin{align}
\label{eq:allgauss2}
A_\theta[x]=& \cos(\theta) A_{q}[x]+ \sin(\theta)A_{p}[x]\nonumber\\
=&\sum_{n=0}^{d_p-1} e^{i\theta n}\psi_n(x)A[n]
\end{align}
where $\psi_n(x)$ are the energy eigenfunctions of the harmonic oscillator. We assume that the physical dimension is finite and discuss the required changes due to an infinite physical dimension afterwards.

We now argue that Requirement \ref{req:approximation} cannot be fulfilled in the present situation:
\begin{observation}[Impossibility of control]
\label{obs:nocontrol}
Let $A_\theta[x]$ be as in (\ref{eq:allgauss2}) with finite $d_p$. Then, there exist some constants $C>0$, $\lambda>0$, and $\eps_0>0$ such that for all $U\in\mathbb{SU}_2$, $\theta\in[0,\pi]$, and all $0<\eps\le\eps_0$  
\begin{equation}
\label{eq:impcontrol}
\PP_\theta(\|\bar{A}_\theta[x]-U\|\le\eps)<C\eps^\lambda,
\end{equation}
where $\PP_\theta$ denotes the probability for the random variable $x$ which corresponds to a quadrature measurement with angle $\theta$. 
\end{observation}
If this is true, the expected number of tries to implement any unitary is bounded as $\EE(n_\eps)>C^{-1}(1/\eps)^{\lambda}$ for $0<\eps\le\eps_0$, and Requirement \ref{req:approximation} is not fulfilled. The proof is given in the appendix.

This result, which can be summarized as ``no quantum computation with Gaussian measurements only'' is complementary to the one reported in Ref.\ \cite{nogo} which can, in turn, be summarized as ``no quantum computing with Gaussian states only''. These results apply only to the model where all operations are noiseless and error-correction is not performed. Together, this means, both non-Gaussian states and non-Gaussian measurements are simultaneously necessary for continuous variable MBQC. However, the two limitations are of very different nature. The reason for the former is that there exists not enough localizable entanglement which means that the quantum information gets destroyed along the wire and transport or teleportation is not possible. The root of the latter limitation is the insufficient possibility to control what quantum gate is performed. 

\subsection{Infinite-dimensional physical systems}

Up to now, we have only considered resource-states where the physical dimension is finite. We demonstrate the difference between states with and without support on a finite number of Fock states with the help of an example where the resource state is described by the matrices
\begin{equation}
A[x]=\frac{1}{\sqrt{\mathcal{N}}}\left(f(x)\id+ig(x)Z\right)
\end{equation}
with $\mathcal{N}=\int\mathrm{d}x\,(|f(x)|^2+|g(x)|^2)$ to ensure the normalization condition 
\begin{equation}
	\int\mathrm{d}x\,A[x]^\dag A[x]=\id.
\end{equation}	
Both $f$ and $g$ are chosen to be real, even functions. Thus, also their Fourier transforms $\tilde{f}$, $\tilde{g}$ are real and the matrices corresponding to $\hat{q}_\theta$ are 
\begin{multline}
A_\theta[x]=\frac{1}{\sqrt{\mathcal{N}}}\bigl((f(x)\cos(\theta)\\
+\tilde{f}(x)\sin(\theta))\id+i(g(x)\cos(\theta)+\tilde{g}(x)\sin(\theta))Z\bigr),
\end{multline}
which are all unitary. If the resource state is finite-dimensional, $f$, $g$, $\tilde{f}$, and $\tilde{g}$ are of the form 
\begin{equation}	
	h(x)={\rm poly}(x)\exp(-x^2/2)
\end{equation}	
and the probability for $\bar{A}_\theta[x]$ to differ too much from the desired unitary is too large. If we allow the resource state to be infinite dimensional, the situation changes. In particular, it is possible that $f(x)={\rm const}$ for $|x|\le c$ for some constant $c>0$. This allows to find a resource state where a $q$-quadrature measurement implements some gate with a finite probability. This is a striking example for a qualitative difference between a proper continuous-variable state and all of its finite dimensional truncations. 

One might now be tempted to use this new insight to devise a MBQC scheme which allows for theoretically efficient single-qubit processing with Gaussian measurements only, but this is impossible. This can be immediately seen by truncating the state according to (\ref{eq:meanphoton}). Then, the above results imply that single qubit MBQC is impossible when the energy per mode stays constant when scaling the system. If, on the other hand, the energy per mode is increased with the length of the computation, this no-go result does not hold any longer. 

The reasons for the existence of such a severe limitation lies in the strong notion of efficiency. If Eq.~(\ref{eq:coupling}) in Requirement \ref{req:approximation} is weakened to (\ref{eq:coupling2}), this restriction is not present anymore, and it is possible to devise a MBQC-scheme based on Gaussian measurements on a non-Gaussian resource. To do so, one can take the resource defined by the matrices (\ref{eq:Aq}). After changing the variable and neglecting global phases, one gets 
\begin{equation}
\bar{A}[x]=S(-p_0q_0)S(-2p_0x)
\end{equation}
with $p(x)=\exp(-x^2)/\sqrt{\pi}$ where  

\begin{eqnarray}
	S(\phi)&=&{\rm diag}(\exp(-i\phi/2),\exp(i\phi/2))\nonumber\\
	&=&\exp(-i\phi/2){\rm diag}(1,\exp(i\phi))
\end{eqnarray}	
is the phase gate. Assume that the target gate is $S(\phi)$. Chose $q_0$ and $p_0$ such that $\theta=-q_0p_0$ and, therefore, $\bar{A}[0]=S(\theta)$. Using $\|S(\alpha)-S(\beta)\|\le|\alpha-\beta|$, one can bound the probability that the distance of the actually implemented gate to the desired one is larger than $\eps$ as
\begin{equation}
\PP(\|\bar{A}[x]-S(\theta)\|>\eps)\le\PP(|2p_0x|>\eps)\le Cp_0\eps
\end{equation}
where $C$ is some constant. This means, it is possible to approximate an arbitrary phase gate in $O(1/\eps)$ steps. Changing the interaction unitary to contain an additional Hadamard gate on the correlation space and using the decomposition detailed below together with the composition law for errors, it is easy to show that any $U\in\mathbb{SU}_2$ can be approximated in $O((1/\eps)^4)$ steps. 

\section{Feasible, efficient CV-MBQC schemes}

In this Section, we present a scheme which allows for efficient CV-MBQC. The limitations discussed above force us to use both non-Gaussian resource states and non-Gaussian measurements. Even though the experimental requirements for its realization are indeed challenging, the scheme does only use primitives which all have already been demonstrated to be feasible in proof-of-principle experiments. 

\subsection{Single qubit operations}
\label{sus:efficiency}
 We choose the interaction unitary in Eq.\ (\ref{eq:interactionunitary}) to be
\begin{equation}
\label{eq:exunitary}
U=\left(H\otimes\id\right)\left(|0\rangle\langle 0|\otimes\exp(-i\theta\hat{n})+|1\rangle\langle 1|\otimes\exp(i\theta\hat{n})\right)
\end{equation}
where $\theta>0$ is a parameter. We initialize the modes in a coherent state vector $|\alpha\rangle$ with $\alpha>0$. Eq.~(\ref{eq:exunitary}) describes a controlled rotation in phase space. The class of measurements corresponding to unitary evolution is given by the displaced photon-counting measurements, i.e., projections onto the state vector $|x,n\rangle:=\hat{D}(x)|n\rangle$ where $x\in\rr$. With the help of (\ref{eq:defalpha}), the applied gates read
\begin{equation}
\label{eq:Axn}
\bar{A}_x[n]=H\left[\begin{array}{cc} e^{-in\phi(x)} & 0 \\ 0 & e^{in\phi(x)}\end{array}\right]
\end{equation}
where
\begin{equation}
\label{eq:phix}
\phi(x)={\rm arctan}\,\left(\frac{\alpha\sin(\theta)}{\alpha\cos(\theta)+x}\right).
\end{equation}
The corresponding probabilities are
\begin{equation}
\label{eq:probx}
p_x(n)=e^{-(\alpha^2+x^2+2x\cos(\theta))}\frac{(\alpha^2+x^2+2x\cos(\theta))^{n}}{n!}
\end{equation}
and fulfill, for every $x$, the normalization condition $\sum_np_x(n)=1$.  

Any single-qubit unitary operation can be efficiently approximated with those operations: Every $U\in\mathbb{SU}_2$ can be written as 
\begin{equation}
\label{eq:su2decomp}
U=S(\phi_1)HS(\phi_2)HS(\phi_3)
\end{equation}
 $H$ is the Hadamard gate, and $\phi_i\in[0,2\pi]$.  Rewriting this as
\begin{equation}
\label{eq:su2decomp2}
U=HS(0)HS(\phi_1)HS(\phi_2)HS(\phi_3),
\end{equation}
we have decomposed every gate into four applications of (\ref{eq:Axn}). Inverting (\ref{eq:phix}) yields
\begin{equation}
x(\phi)=\frac{\alpha\sin(\theta)}{\tan(\phi(x))}-\alpha\cos(\theta).
\end{equation}
To implement $S(\phi)$, one only has to choose $x$ such that $-2\phi(x) n=\phi$ for some small $n$ where $n=0$ and $n=1$ suffice and the former is only needed  for $\phi=0$. For any fixed $\alpha\neq 0$ and $\theta\neq 0$, the maximally necessary shift $|x|$ is bounded and with (\ref{eq:probx}) there exists a lower bound $0<p_0 \le p_x(n)$ for all relevant $x$ and $n$. Using (\ref{eq:su2decomp2}), one can see that the probability to obtain any chosen $U\in\mathbb{SU}_2$ in four steps is lower bounded by $p_0^4$. Thus, the conditions for Observation \ref{obs:randwalkcond} are fulfilled and efficient single-qubit processing is possible. 

\subsection{Coupling of wires}
\label{sub:coulinex}

We now turn to the coupling of two wires. Let the two correlation space qubits be prepared in $|\psi\rangle_{j}=a_{j}|0\rangle+b_{j}|1\rangle$ for $j=1,2$. Then, the state vectors after the interaction with the light mode reads
\begin{equation}
a_{j}|0\rangle|e^{-i\theta}\alpha\rangle+b_{j}|1\rangle|e^{i\theta}\alpha\rangle.
\end{equation}
Displacing now both modes by $\Delta_x=\alpha\cos(\theta)$, rotating the second mode by an angle of $\pi/2$, and setting $\gamma=\alpha \sin(\theta)$, the joint state vector becomes
\begin{multline}
\label{eq:tt1}
a_{1}a_{2}|0,0\rangle|\gamma\rangle|i\gamma\rangle+
a_{1}b_{2}|0,1\rangle|\gamma\rangle|-i\gamma\rangle\\+
b_{1}a_{2}|1,0\rangle|-\gamma\rangle|i\gamma\rangle+
b_{1}b_{2}|1,1\rangle|-\gamma\rangle|-i\gamma\rangle.
\end{multline}
A balanced beam-splitter transforms two coherent states as
\begin{equation}
\label{eq:beamsplitterA}
|\alpha\rangle|\beta\rangle\mapsto|\frac{\alpha+\beta}{\sqrt{2}}\rangle|\frac{\alpha-\beta}{\sqrt{2}}\rangle.
\end{equation}
Applying this to (\ref{eq:tt1}) yields
\begin{align}
\label{eq:tt2}
&a_{1}a_{2}|0,0\rangle|\frac{\gamma}{\sqrt{2}}(1+i)\rangle|\frac{\gamma}{\sqrt{2}}(1-i)\rangle\nonumber\\+
&a_{1}b_{2}|0,1\rangle|\frac{\gamma}{\sqrt{2}}(1-i)\rangle|\frac{\gamma}{\sqrt{2}}(1+i)\rangle\nonumber\\+
&b_{1}a_{2}|1,0\rangle|\frac{\gamma}{\sqrt{2}}(-1+i)\rangle|\frac{\gamma}{\sqrt{2}}(-1-i)\rangle\nonumber\\+
&b_{1}b_{2}|1,1\rangle|\frac{\gamma}{\sqrt{2}}(-1-i)\rangle|\frac{\gamma}{\sqrt{2}}(-1+i)\rangle.
\end{align}
The matrices corresponding to a photon-counting measurement in both modes with results $n_1$ and $n_2$ respectively read
\begin{align}
A[n_1,n_1]=&{\rm diag}\,(e^{-\frac{i\pi n_1}{4}},e^{-\frac{i7\pi n_1}{4}},e^{-\frac{i3\pi n_1}{4}},e^{-\frac{i5\pi n_1}{4}})\nonumber\\
&\times{\rm diag}\,(e^{-\frac{i7\pi n_2}{4}},e^{-\frac{i\pi n_2}{4}},e^{-\frac{i5\pi n_2}{4}},e^{-\frac{i3\pi n_2}{4}})
\end{align}
which is, up to a local unitary, a controlled phase gate, i.e.,
\begin{multline}
\label{eq:An1n2}
A[n_1,n_2]={\rm diag}\,(e^{-\frac{i\pi(n_1+7n_2)}{4}},e^{-\frac{i\pi(3n_1+5n_2)}{4}})\\
\otimes{\rm diag}\,(1,e^{-\frac{i\pi(6n_1-6n_2)}{4}})\\\times{\rm diag}\,(1,1,1,e^{-i\pi(-n_1+n_2)}).
\end{multline}
If $n_1+n_2$ is even, $A[n_1,n_2]$ is not entangling, and the local operations can be efficiently undone as described above. If $n_1+n_2$ is odd, a $C_Z$-gate is implemented up to local correction. As this situation occurs with finite probability for any choice of $\alpha$ and $\theta$, this means that Requirement \ref{req:coupling} is fulfilled. 

It remains to show that initialization and read-out are possible. To measure in the computational basis, we shift the mode after the interaction by $\Delta=\alpha(\cos(\theta)-i\sin(\theta))$ which turns the joint state of qubit and mode to
\begin{equation}
a|0\rangle|0\rangle+b|1\rangle|2i\alpha\sin(\theta)\rangle.
\end{equation}
Counting the photons leads to a matrix fulfilling the conditions of Requirement \ref{req:iniandreadout}. Note that there is some asymmetry: If the counted number of photons is larger than one, the qubit is in $|1\rangle$. If the result is zero, one has to undo the gate $H{\rm diag}\,(1,-i)$ and repeat the procedure. Due to the gap between $1$ and $|\langle 0|2i\alpha\sin(\theta)\rangle|$, this is efficiently possible. This concludes the proof that the proposed scheme is efficient. 

\begin{figure}
\begin{center}
\includegraphics{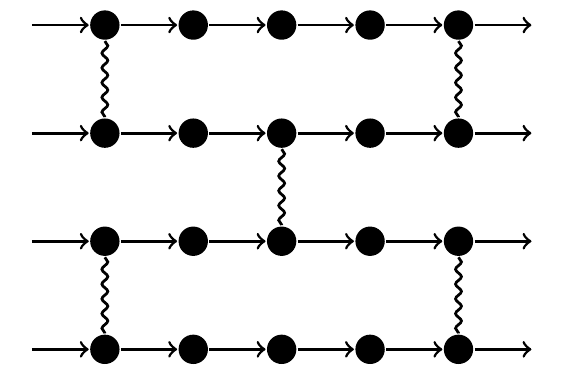}
\end{center}
\caption{\label{fig:realMBQC} Non-adaptive coupling of four quantum wires where the wiggly lines denote the probabilistic measurement-based $C_Z$ gate. By scaling the distance between the single couplings polylogarithmically in the length of the computation, the total probability of failure can be made arbitrarily small.}
\end{figure}

As already mentioned, the coupling scheme discussed so far is not completely in the MBQC-paradigm because it involves the adaptive coupling of sites. However, this is not necessary: Consider two wires which are coupled every $k+1$ sites by the beam-splitter described by Eq.\ (\ref{eq:beamsplitterA}). Because $C_Z^2=\id$, two consecutive couplings can be undone by appropriately choosing local operations between them. From Eq.\ (\ref{eq:probx}) it follows that for every block of four sites, there exists a probability $p_0>0$ that the desired gate is realized and probability of failure in at least one of the wires is upper bounded by $2(1-p_0)^{k/4}$. Let $n$ be total number of gates, the total probability of failure is bounded by 
\begin{equation}
p_{\rm failure}\le 2n(1-p_0)^{k/4}\le c_1ne^{-c_2k}
\end{equation} 
for suitable constants $c_1$ and $c_2$. Thus, for a fixed probability of success, $k$ has only to grow as $k=O(\polylog(n))$, which means that the need for two-site measurements can be removed for the price of a polylog overhead. For a quantum circuit consisting of more than two wires, as depicted in Figure \ref{fig:realMBQC}, every wire should be alternately coupled to its left and right neighbor.  When the circuits consists of $m$ quantum wire with all have a length $n$, the respective overhead behaves as $k=O(\polylog(nm))$.

\subsection{Errors}
\label{sus:errors}
Even if a model fulfills the theoretical requirements it will still be affected by noise. Thus, error correction will be necessary to build a scalable quantum computer. In the scheme presented above, one plausible
dominant source of errors is reflecting the finite efficiency of the photon-counting measurement 
needed to perform both single and two qubit gates. If photon loss happens, a ``wrong'' operation is applied in the two-dimensional correlation space. When the rate of these qubit errors is low enough, techniques of error correction and fault-tolerant quantum computing may be applied to make a computation with unbounded length possible at the expense of a polylogarithmic overhead \cite{faulttolerantnielsen,faulttolerantraussendorf}. This contrasts with the errors stemming from the use of finitely squeezed states in the scheme based on Gaussian cluster states for which no method of error correction exists yet to achieve fault tolerance. 

Beside of the obvious errors stemming from decoherence and imperfections in the measurement procedure, the presented continuous variable schemes have an important additional source of errors. For example, when performing a shifted photon counting, uncertainty in the parameter of the shift operator $D(x)$ results in uncertainty of the applied operator. We now investigate how those errors scale with the length of the computation. We restrict ourselves to the single-qubit case while noting that the error analysis for multiple qubits is completely analogous. Consider the situation where the target gate reads $U=\prod_{i=1}^nU_i$ while the gate actually applied is $\bar{A}=\prod_{i=1}^n\bar{A}_i$. We assume that the error induced by a single step is bounded, i.e., for all $i$, 
\begin{equation}
	\|\bar{A}_i-U_i\|\le\eps.
\end{equation}	
 If all $\bar{A}_i$ are also unitary, the total error grows linearly in $n$ and can be bounded as $\|\bar{A}-U\|\le n\eps$. If this is not the case, an tight upper bound to the error is 
\begin{equation}
	\|\bar{A}-U\|\le n\eps(1+\eps)^n,
\end{equation}
 i.e., it grows, in the worst case, exponentially in $n$. In the MBQC-scheme based on Gaussian cluster states as discussed, e.g., in Refs.\ \cite{gaussiancluster1,gaussiancluster2,gaussiancluster3,gaussiancluster4}, the implemented gates are not unitary for physical resource states which is the reason for the exponential decay of the transport fidelity \cite{nogo}.

\section{Conclusion}
In this work, we have investigated the possibilities provided by measurement-based quantum computing with continuous-variable resources. After introducing a framework allowing for the description of a huge class of possible resource states, we have clarified which conditions are necessary for a CV-MBQC scheme to be viable. Those conditions lead to new limitations. Especially, we have shown that, without the use of not yet developed continuous-variable error correction, scalable quantum computing with Gaussian measurements alone is impossible, even if the resource state is non-Gaussian, complementing the prior result which prohibits MBQC with Gaussian resources, even if the measurements are non-Gaussian. As the second main result, we gave an explicit example of an efficient MBQC scheme where the resource could be created by a simple interaction between light modes in coherent states and some qubit degrees of freedom. Processing is then performed by shifted photon counting while entangling two-qubit gates rely on simple beam splitters. An analysis of errors highlighted the qualitative differences of the two major kinds of errors. 

Two major conclusions can be drawn from our findings: First, continuous-variable measurement based quantum computing 
is an extraordinarily difficult enterprise. Second, when non-Gaussian resource states are combined with non-Gaussian measurements, efficient schemes do exist, and in this work, we introduce such first strictly efficient schemes. While this may be disappointing when aiming at a scalable optical quantum computer, one does need to keep in mind, that our statements about efficiency only address the asymptotic behavior. Thus, the limitation do not at all rule out the usefulness of much simpler protocols in situations where only a few modes are used. Such situations include quantum repeaters and hybrid entanglement-distribution protocols for applications particularly in quantum cryptography. Hence, for such purposes, both the purely Gaussian setting as well as those non-Gaussian schemes that asymptotically do not give rise to universal quantum computing can well be feasible as elementary tools for entanglement distillation schemes.
It is the hope that the framework established here fosters
further such investigations.

\section{Acknowledgments.}

We would like to thank the EU (Qessence, Compas, Minos), the BMBF (QuOReP), and the EURYI for support.

\begin{appendix}

\section{Proof of Observation \ref{obs:nocontrol}}
To find an upper bound to $\PP(\|\bar{A}_\theta[x]-U\|\le\eps)$, we first lower bound the probability that $|x|>x_0$ for some suitably chosen $x_0$. Then, for $|x|\le x_0$, we upper bound $\|\bar{A}_\theta[x]-U\|$ by a polynomial in $x$ and use that the probability for a polynomial to be larger than $\eps$ cannot grow too fast for small $\eps$. 

We calculate
\begin{equation}
\|\bar{A}_\theta[x]-U\|\ge\|\frac{A_\theta[x]}{\sqrt{p_\theta(x)}}-U\|_2/\sqrt{2}
\end{equation}
where $\|O\|_2=\sqrt{\sum_{i,j}|O_{ij}|^2}$ denotes the Frobenius norm. We now use (\ref{eq:allgauss2}) and the fact that 
\begin{equation}
\psi_n(x)=\frac{1}{\sqrt{2^nn!\sqrt{\pi}}}H_n(x)e^{-x^2/2},
\end{equation}
where $H_n(x)$ denotes the $n$th Hermite polynomial. We also use $0\le\sqrt{x}\le x+1$ and that $p_\theta(x)$ is upper-bounded by a constant independent of $\theta$. This allows us to find non-negative polynomials $f_{U,\theta}$ and $g_{U,\theta}$, where the coefficients are continuous functions of $U$ and $\theta$ with $\|\bar{A}_\theta[x]-U\|\ge f_{U,\theta}(x)/g_{U,\theta}(x)$. As the probability density $p_\theta(x)$ is of the form $p_\theta(x)=\poly(x)\exp(-x^2)$ we get
\begin{equation}
\label{eq:boundlargedev}
\PP_\theta\left(|x|>C_1\log\left(\frac{C_2}{\eps}\right)\right)\le\eps
\end{equation}
for some suitable $C_1>0$, $C_2>0$. Now, we assume that $|x|\le C_1\log\left(C_2/\eps\right)$. In this case, $g_{U,\theta}$ can be upper bounded by $C_3\left(C_1\log\left(C_2/\eps\right)\right)^r$ for some $C_3>0$ and some even $r$. Note that the maximal degree of all polynomials only depends on $d_p$. Assume that $f_{U,\theta}$ has $M$ zeros $x_1,\ldots,x_M$. Around any of these zeros, $f_{U,\theta}$ can be lower bounded by $C_4(x-x_i)^s$ for some $C_4>0$ and some even $s$. Using this and (\ref{eq:boundlargedev}), we chose some $\eps_0$ such that
\begin{align}
&\PP_\theta\left(\|\bar{A}_\theta[x]-U\|\le \eps\right)\le \eps\\
&+\sum_{i=1}^M\PP_\theta\Biggl(C_4(x-x_i)^s\le C_3\left(C_1\log\left(\frac{C_2}{\eps}\right)\right)^r\eps\Biggr).\label{eq:app0}
\end{align}
for all $\eps\le\eps_0$. Because $p_\theta(x)$ is upper bounded, there exist a constant $C_5$ such that 
\begin{equation}
\label{eq:app1}
\PP((x-x_i)^s\le\eps)\le C_5\eps^{1/s}. 
\end{equation}
for all $\eps\le\eps_0$. Inserting (\ref{eq:app1}) into (\ref{eq:app0}), one obtains (\ref{eq:impcontrol}). As all constants depend on $U$ and $\theta$ in a continuous way, and $U$ and $\theta$ are both taken from compact domains, there exist a set of constants such that (\ref{eq:impcontrol}) is simultaneously true for all $U$ and $\theta$. Thus, Observation \ref{obs:nocontrol} holds.

\end{appendix}


\begin{thebibliography}{10}

\bibitem{clusterbriegel}
	R.\ Raussendorf and H.~J.\ Briegel,
	 Phys.\ Rev.\ Lett.\ {\bf {\bf 86}}, 5188 (2001).

\bibitem{clusterlong}
	R.\ Raussendorf and H.~J.\ Briegel,
	Quant.\ Inf.\ Comp.\ {\bf 6}, 433 (2002).

\bibitem{prl}
	D.\ Gross and J.\ Eisert,
	 Phys.\ Rev.\ Lett.\ {\bf  98}, 220503 (2007).
	 
\bibitem{pra}	 
	 D.\ Gross, J.\ Eisert, N.\ Schuch, and D.\ Perez-Garcia, 
	Phys.\ Rev.\ A {\bf 76}, 052315 (2007).

\bibitem{cvreview1}	
	J.\ Eisert and M.~B.\ Plenio,
	Int.\ J.\ Quant.\ Inf.\ {\bf 1}, 479 (2003).

\bibitem{cvreview2}
	S.~L.\ Braunstein and P.\ van Loock,
	 Rev.\ Mod.\ Ph/ys.\ {\bf 77}, 513 (2005).	

\bibitem{ng1}
	J.\ Eisert, S.\ Scheel, and M.~B.\ Plenio,
	 Phys.\ Rev.\ Lett.\ {\bf 89}, 137903 (2002).

\bibitem{ng2}
	J.\ Fiur\'{a}\v{s}ek,
	 Phys.\ Rev.\ Lett.\ {\bf 89}, 137904 (2002).

\bibitem{ng3}
	G.\ Giedke and J.~I.\ Cirac,
	 Phys.\ Rev.\ A {\bf 66}, 032316 (2002).

\bibitem{cerf}
	J.\ Niset, J.\ Fiur\'{a}\v{s}ek, and N.~J.\ Cerf,
	Phys.\ Rev.\ Lett. {\bf 102}, 120501 (2009).
	
\bibitem{efficientsimulation}
	S.~D.\ Bartlett, B.~C.\ Sanders, S.~L.\ Braunstein, and K.\ Nemoto,
	Phys.\ Rev.\ Lett. {\bf 88}, 097904 (2002).

\bibitem{gaussiancluster1}
	N.~C.\ Menicucci, 
	P.\ van Loock, M.\ Gu, 
	C.\ Weedbrook, T.~C.\ Ralph, and M.~A.\ Nielsen,
        Phys.\ Rev.\ Lett.\ {\bf {\bf 97}}, 110501 (2006).

\bibitem{gaussiancluster2}
	N.~C.\ Menicucci, S.~T.\ Flammia, and O.\ Pfister,
	Phys.\ Rev.\ Lett.\ {\bf 101}, 130501 (2008).

\bibitem{gaussiancluster3}
	S.~T.\ Flammia, N.~C.\ Menicucci, and O.\ Pfister,
	J.\ Phys.\ B {\bf 42}, 114009 (2009).

\bibitem{gaussiancluster4}
	R.\ Ukai, J.-I.\ Yoshikawa, N.\ Iwata, P.\ van Loock, and A.\ Furusawa,
	Phys.\ Rev.\ A {\bf 81}, 032315 (2010). 

\bibitem{gaussiancluster5}
	M.\ Gu, C.\ Weedbrook, N.\ C.\ Menicucci, T.\ C.\ Ralph, and P.\ van Loock,
	Phys.\ Rev.\ A {\bf 79}, 062318 (2009).

\bibitem{nogo}
	M.\ Ohliger, K.\ Kieling, and J.\ Eisert,
	Phys.\ Rev.\ A {\bf 82}, 042336 (2010).

\bibitem{faulttolerantnielsen}
	M.~A.\ Nielsen and C.~M.\  Dawson,
	Phys.\ Rev.\ A {\bf 71}, 042323 (2005).

\bibitem{faulttolerantraussendorf}
	R.\ Raussendorf and J.\ Harrington,
	Phys.\ Rev.\ Lett.\ {\bf 98}, 190504 (2007).
	
\bibitem{cats}
	T.~C.\ Ralph, A.\ Gilchrist, G.~J.\ Milburn, W.~J.\ Munro, and S.\ Glancy,
    Phys.\ Rev.\ A {\bf 68}, 042319 (2003).

\bibitem{hybrid1}
	T.~P.\ Spiller, K.\ Nemoto, S.~L.\ Braunstein, W.~J.\ Munro, P.\ van\ Loock, and G.~J.\ Milburn,
	New.\ J.\ Phys {\bf 8}, 30 (2006).

\bibitem{hybrid2}
	P.\ van Loock, W.\ J.\ Munro, K.\ Nemoto, T.\ P.\ Spiller, T.\ D.\ Ladd, S.\ L.\ Braunstein, 
	  and G.\ J.\ Milburn,
	Phys.\ Rev.\ A {\bf 78}, 022303 (2008).

\bibitem{schleich}
	W.\ P.\ Schleich, 
	\textit{Quantum optics in phase space}, 
	Wiley-VCH (2001).

\bibitem{crosskerr}
	P.\ Grangier, J.~A.\ Levenson, and J.-P.\ Poizat,
	Nature {\bf 396}, 537 (1998).


\bibitem{multiplexing}
	D.\ Achilles, C.\ Silberhorn, C.\ Sliwa, 
	 K.\ Banaszek, I.\ A.\ Walmsley, M.\ J.\ Fitch, B.\ C.\ Jacobs, T.\ B.\ Pittman, 
	 and J.\ D.\ Franson,
	J.\ Mod.\ Opt.\ {\bf 51}, 1499 (2004).

\bibitem{nanowires}
	A.\ Divochiy, F.\ Marsili, D.\ Bitauld, A.\ Gaggero, R.\ Leoni, F.\ Mattioli, A.\ Korneev, V.\ Seleznev, N.\ Kaurova, O.\ 	
	   Minaeva, G.\ Goltsman, K.G.\ Lagoudakis, M.\ Benkhaoul, F.\ Levy, and A.\ Fiore,
	Nature Photonics {\bf 2}, 302 (2008). 

\bibitem{singleatom1}
	S.\ Rebic, S.~M.\ Tan, A.~S.\ Parkins, and D.~F.\ Walls,
	J.\ Opt.\ B.\  {\bf 1}, 490 (1999).

\bibitem{singleatom2}
	S.\ Parkins,
	Nature {\bf 465}, 699 (2010).

\bibitem{webs}
	D.\ Gross and J.\ Eisert,
	Phys.\ Rev.\ A {\bf 82}, 040303(R) (2010).

\bibitem{mpstheory1}
	M.\ Fannes, B.\ Nachtergaele, and R.~F.\ Werner,
	 {Comm.\ Math.\ Phys.} {\bf {\bf 144}}, {443} ({1992}).

\bibitem{mpstheory2}
	D.\ Perez-Garcia, F.\ Verstraete, M.~M.\ Wolf, and J.~I.\ Cirac,
	Quant.\ Inf.\ Comp {\bf {\bf 7}}, 401 (2007).

\bibitem{cvcs}
	M.\ Ohliger, V.\ Nesme, D.\ Gross, and J.\ Eisert,
	arXiv:1111.0853.

\bibitem{sequentialpreparation}
	C.\ Sch\"on, K.\ Hammerer, M.~M.\ Wolf, J.~I.\ Cirac, and E.\ Solano,
    	Phys.\ Rev.\ A {\bf 75}, 032311 (2007).

\bibitem{groundstates1}
	G.\ K.\ Brennen and A.\ Miyake,
	Phys.\ Rev.\ Lett.\ {\bf 101}, 010502 (2008).

\bibitem{groundstates2}
	X.\ Chen, B.\ Zeng, Z.\ Gu, B.\ Yoshida, and I.\ L.\ Chuang,
	Phys.\ Rev.\ Lett.\ {\bf 102}, 220501 (2009).

\bibitem{cMPS1}
	F.\ Verstraete and J.\ I.\ Cirac,
	Phys.\ Rev.\ Lett.\ {\bf 104}, 190405 (2010).

\bibitem{cMPS2}
	T.\ J.\ Osborne, J.\ Eisert, and F.\ Verstraete,
	Phys.\ Rev.\ Lett.\ {\bf 105}, 260401 (2010).


\end{thebibliography}
\end{document}